\documentstyle[charm2000]{article}
\def\etal{{\it et al.}}

\def\fm3{fm$^3$}
\def\fmm3{fm$^{-3}$}

\def \ba88{{\it Particles and Fields 3} (Proceedings of the 1988 Banff Summer
Institute on Particles and Fields), edited by A. N. Kamal and F. C. Khanna
(World Scientific, Singapore, 1989)}

\def \be87{{\it Proceedings of the Workshop on High Sensitivity Beauty
Physics at Fermilab,} Fermilab, Nov. 11-14, 1987, edited by A. J. Slaughter,
N. Lockyer, and M. Schmidt (Fermilab, Batavia, IL, 1988)}

\def \cp89{{\it CP Violation,} edited by C. Jarlskog (World Scientific,
Singapore, 1989)}
\def \dpf91{{\it The Vancouver Meeting - Particles and Fields '91}
(Division of Particles and Fields Meeting, American Physical Society,
Vancouver, Canada, Aug.~18-22, 1991), ed. by D. Axen, D. Bryman, and M. Comyn
(World Scientific, Singapore, 1992)}

\def \hb87{{\it Proceeding of the 1987 International Symposium on Lepton and
Photon Interactions at High Energies,} Hamburg, 1987, ed. by W. Bartel
and R. R\"uckl (Nucl.~Phys.~B, Proc. Suppl., vol. 3) (North-Holland,
Amsterdam, 1988)}

\def \ite{{\it et al.}}

\def \ky85{{\it Proceedings of the International Symposium on Lepton and
Photon Interactions at High Energy,} Kyoto, Aug.~19-24, 1985, edited by M.
Konuma and K. Takahashi (Kyoto Univ., Kyoto, 1985)}
\def \lat90{{\it Results and Perspectives in Particle Physics} (Proceedings of
Les Rencontres de Physique de la Vallee d'Aoste [4th], La Thuile, Italy, Mar.
18-24, 1990), edited by M. Greco (Editions Fronti\`eres, Gif-Sur-Yvette,
France,
1991)}
\def \lg91{International Symposium on Lepton and Photon Interactions, Geneva,
Switzerland, July, 1991}
\def \lkl87{{\it Selected Topics in Electroweak Interactions} (Proceedings of
the Second Lake Louise Institute on New Frontiers in Particle Physics, 15 --
21 February, 1987), edited by J. M. Cameron \ite~(World Scientific, Singapore,
1987)}

\def \np#1#2#3{{\it Nucl. Phys.} {\bf#1} (#3) #2}
\def \oxf65{{\it Proceedings of the Oxford International Conference on
Elementary Particles} 19/25 Sept.~1965, ed.~by T. R. Walsh (Chilton, Rutherford
High Energy Laboratory, 1966)}

\def \pl#1#2#3{{\it Phys. Lett.} {\bf#1} (#3) #2}

\def \pr#1#2#3{{\it Phys. Rev.} {\bf#1} (#3) #2}

\def \prl#1#2#3{{\it Phys. Rev. Lett.} {\bf#1} (#3) #2}

\def \si90{25th International Conference on High Energy Physics, Singapore,
Aug. 2-8, 1990, Proceedings edited by K. K. Phua and Y. Yamaguchi (World
Scientific, Teaneck, N. J., 1991)}
\def \slac75{{\it Proceedings of the 1975 International Symposium on
Lepton and Photon Interactions at High Energies,} Stanford University, Aug.
21-27, 1975, edited by W. T. Kirk (SLAC, Stanford, CA, 1975)}
\def \slc87{{\it Proceedings of the Salt Lake City Meeting} (Division of
Particles and Fields, American Physical Society, Salt Lake City, Utah, 1987),
ed. by C. DeTar and J. S. Ball (World Scientific, Singapore, 1987)}
\def \smass82{{\it Proceedings of the 1982 DPF Summer Study on Elementary
Particle Physics and Future Facilities}, Snowmass, Colorado, edited by R.
Donaldson, R. Gustafson, and F. Paige (World Scientific, Singapore, 1982)}
\def \smass90{{\it Research Directions for the Decade} (Proceedings of the
1990 DPF Snowmass Workshop), edited by E. L. Berger (World Scientific,
Singapore, 1991)}

\def \tasi90{{\it Testing the Standard Model} (Proceedings of the 1990
Theoretical Advanced Study Institute in Elementary Particle Physics),
edited by M. Cveti\v{c} and P. Langacker (World Scientific, Singapore, 1991)}

\begin{document}
\title{Pentaquark Search with Energetic Hadron Beams}
\author{M. A. Moinester,$^{(1)}$
D. Ashery,$^{(1)}$
L. G. Landsberg,$^{(2)}$
H. J. Lipkin$^{(1,3)}$\\
         }  
\maketitle
\vspace{-3.6in}
\leftline{TAUP-2179-94}
\leftline{July 1994}
\centerline{Bulletin Board HEP-PH@xxx.lanl.gov --- 9407319}
\vspace{-0.6in}
\rightline{Presented at CHARM2000 Workshop}
\rightline{Fermilab, June 7 -- 9, 1994}
\vspace{1.3in}
\begin{center}
$^{(1)}$Raymond and Beverly Sackler Faculty of Exact Sciences,
School of Physics\\
Tel-Aviv University, 69978 Ramat Aviv, Israel\\
$^{(2)}$Institute for High Energy Physics, 142284 Protvino, Russia\\
$^{(3)}$Department of Nuclear Physics, Weizmann Institute of Science,
76100 Rehovot, Israel\\
\end{center}
\begin{abstract}
 The strange-anticharmed Pentaquark is a $uud\bar{c}s$ or $udd\bar{c}s$
five-quark baryon that is expected to be either a narrow resonance, or
possibly even stable against strong decay. We describe this hyperon here;
its structure, binding energy and lifetime, resonance width, production
mechanisms and decay modes. We estimate production cross sections,
techniques to reduce backgrounds in search experiments, and how to optimize
experiments to observe it. Possibilities for enhancing the signal over
background in Pentaquark searches are investigated by examining predictions
for detailed momentum and angular distributions in multiparticle final
states. General model-independent predictions are presented as well as
those from two models: a loosely bound $D_{s}^-N$ "deuteron" and a
strongly-bound five-quark model. Fermilab E791 data, currently being
analysed, may be marginal for showing definitive signals. Future
experiments with more than $10^5$ reconstructed charmed baryon events
should have sensitivity to determine whether or not the Pentaquark exists.
\end{abstract}

\section{Introduction}

Ordinary hadrons are mesons or baryons, whose quantum numbers can be
described by quark-antiquark or three-quark configurations. Unusual hadrons
that do not fit this picture would constitute new forms of hadronic matter
- exotic hadrons. Such hadrons may have significant multiquark
configurations such as $qq\bar{q}\bar{q}$ and $qqqq\bar{q}$. Exotic hadrons
can have anomalous quantum numbers not accessible to a three-quark or
quark-antiquark structures (open exotic states) or even usual quantum
numbers ( cryptoexotic states). Cryptoexotic hadrons can be identified only
by their unusual dynamical properties (anomalously narrow decay widths,
anomalous decay branching ratios, etc.). The discovery of exotic hadrons
would have far-reaching consequences for quantum chromodynamics, for the
concept of confinement, and for specific models of hadron structure
(lattice, string and bag models). Detailed discussions of exotic hadron
physics can be found in recent reviews \cite {lgl}.

We consider here possible exotic hadronic states with heavy quarks (c, b),
which contain quarks with four different flavors (e.g. u, d, s, c). Their
properties follow from the general hypothesis of "flavor antisymmetry"
\cite {fa}, by which quark systems characterized by the maximum possible
antisymmetry of quark flavors (both quarks and antiquarks) are the most
strongly bound. For instance, this means that that the $u\bar{u}d\bar{s}$
system would be more bound than the $uu\bar{d}\bar{s}$ one, etc.

   Jaffe \cite {jaf} predicted in this spirit that for dibaryons with six
light quarks, the most bound is the Hexaquark H = [u,u,d,d,s,s]
combination, for which not more than two quarks are in states with
identical flavors. Lipkin \cite {lip} and Gignoux \etal ~ \cite {gig}
showed that 5-quark "anticharmed" baryons (Pentaquarks) of the P$^0$ =
[$uud\bar{c}s$] and P$^-$ = [$udd\bar{c}s$] type, or analogous
"anti-beauty" baryons, are most bound in the 5-quark sector. There are also
predictions \cite {zr} for the most bound tetraquark exotic meson, the
$\tilde{F}_s$=[$cs\bar{u}\bar{d}$].

\section{Binding Energy of the Pentaquark}

Some of these exotic states with heavy quarks may be bound. The masses
would be below the threshold for strong decays (i.e.,
M(P$^0$)~$<$~M($D_{s}^{-}$)~$+$~M($p$)). Such quasi-stable bound states
would decay only via weak interactions, with typical weak decay lifetimes.
Resonant states with masses above the strong decay threshold would decay
strongly. In the present work, we focus on experimental searches for the
Pentaquark, both bound and resonant varieties.

The binding potential of a system is given by the difference between the
Color Hyperfine CH interaction in the system and in the lightest
color-singlet combination of quarks into which it can be decomposed. The
wave function of the H may be written as:
\begin{equation}
 \Psi _{H} = \alpha_{1}  \Psi _{6q} + \beta_{1}
 \Psi _{(\Lambda\Lambda)} + \gamma_{1}  \Psi _{(\Sigma^-\Sigma^+)}
+ \delta_{1}  \Psi _{(\Xi^-p)}.
\end{equation}
The lightest color singlet combination is the $\Lambda\Lambda$ system at
2231 MeV. The CH contribution to the binding energy of the H is about
150~MeV, in simple models of the CH interaction. Similarly, the $P^{0}$ and
P$^-$ wave functions can be written as:
\begin{equation}
 \Psi _{P^0} = \alpha_{2}  \Psi _{5q} + \beta_{2}
 \Psi _{(D_{s}^-p)} + \gamma_{2}  \Psi _{(\Sigma^{+}D^{-})}
+ \delta_{2}  \Psi _{(\Lambda\bar{D}^{0})},
\end{equation}
\begin{equation}
 \Psi _{P^-} = \alpha_{3}  \Psi _{5q} + \beta_{3}
 \Psi _{(D_{s}^-n)} + \gamma_{3}  \Psi _{(\Sigma^{-}\bar{D^{0}})}
+ \delta_{3}  \Psi _{(\Lambda D^{-})}.
\end{equation}
Here the lightest color singlet is the $D_{s}^-N$ system at 2907 MeV. The
CH contribution to the mass splitting M($D_{s}^-p$) - M(P$^0$) is the same
as for the H particle, again in simple models of the color hyperfine
interaction. The anti-Pentaquarks are defined in a similar way and,
in general, whatever will be said about the Pentaquarks will also hold
true for the charge-conjugate particles.

The calculations of ref. \cite {fle} account for the $SU(3)_{F}$ breaking.
It was shown that as the symmetry breaking increases, the P always retains
a larger binding potential than the H and that the binding can be several
tens of MeV. The total binding energy includes the internal kinetic energy.
Because the c quark is massive, the kinetic energy in the P is smaller than
in the H by about 15 MeV. This improves the prospects of the P to be bound.

More recently, Takeuchi, Nussinov and Kubodera \cite {tnk} studied the
effects on the Pentaquark and Hexaquark systems of instanton induced
repulsive interactions for three quarks in flavor antisymmetric states.
They claim in this framework that both Pentaquark and Hexaquark are not
likely to be bound. Also, Zouzou and Richard \cite {zr} reconsidered
previous bag model calculations for the tetraquark and pentaquark. Their
new calculation has weaker chromomagnetic attractions at short distances
and a larger bag radius for multiquark states compared to ordinary hadrons.
They find that the Pentaquark is unbound by 80 MeV, while the $\tilde{F}$
tetraquark is unbound by 230 MeV. Similar conclusions for the P and H were
given by Fleck \etal ~ \cite{fle}. Riska and Scoccola \cite {rs} recently
described the Pentaquark in a soliton model, using different parameter
sets. One set gives a bound state, while another gives a near threshold
resonance. Considering all the uncertainties in knowing the Pentaquark
binding energy, our experimental approach is to search for both strongly
and weakly bound Pentaquarks, as well as unbound Pentaquark resonances.

A very weakly bound $D_{s}^-$p deuteron-size bound state just below
threshold with a structure very different from that of the strongly bound
proton size Pentaquark might still be consistent with these recent
calculations, considering all the model uncertainties. The $D_{s}^-$p
system does not have Pauli blocking and repulsive quark exchange
interactions which arise in all hadron-hadron systems where quarks of the
same flavor appear in both hadrons. Thus, even a comparatively weak short
range interaction could produce a relatively large size bound state
analogous to the deuteron, with a long $D_{s}^-$p tail in its wave function
and a good coupling to the $D_{s}^-$p system. The attraction is due to a
short range interaction, not long-range one-pion exchange. This long
attractive tail will also assist in the production mechanism. Because in
the Pentaquark, unlike the deuteron, there is no short range repulsion, its
structure at short distances will be quite different from that of the
deuteron. This component too has it's influence on the production
mechanism. These issues are discussed in subsection 4.2. The deuteron-like
state will be stable against strong and electromagnetic decays. Since the
$D_{s}^-$p pair is some 50-75 MeV lower mass than other meson-baryon
cluster components in the Pentaquark, it will be the dominant component in
a weakly bound deuteron-like Pentaquark.

\section{Structure and Decay Modes of the Pentaquark}

   There are different possibilities for the internal structure of
observable (not very broad) exotic hadrons. They can be bound states or
near threshold resonance structures of known color singlet sub-systems
($\Lambda \Lambda$ for the H \cite {mm} or $D_{s}^{-}p$ for the P$^0$). But
they can have more complicated internal color structure; such as baryons
with color octet and sextet bonds [$(qqq)_{8c} \times (q\bar{q})_{8c}$] and
[$(qq\bar{q})_{\bar{6}c} \times (qq)_{6c}$] (see ref. \cite{A}). We
designate all such structures as direct five quark configurations. If color
substructures are separated in space by centrifugal barriers, then exotic
hadron resonances can have not very large or even anomalously narrow decay
widths, because of complicated quark rearrangements in the decay processes.
If these exotic hadrons are bound strongly, they can be quasistable, with
only weak decays.

The wave function of the Pentaquark contains two-particle cluster
components, each corresponding to a pair of known color singlet particles;
and also a direct five quark [non-cluster] component. The Pentaquark
production mechanism and its decay modes depend on these components. The
$P^{\circ}$ can be formed for  example by the coalescence of $pD^{-}_{s},
\Lambda \overline{D}^{\circ}, p D_{s}^{*-}, \Sigma^{+} D^{-} +
\Sigma^{\circ} \overline{D}^{\circ}, \Lambda \overline{D}^{*\circ},
\Sigma^+ D^{*-} + \Sigma^{\circ} \overline{D}^{*\circ}$; or by a one-step
hadronization process. Let us consider three color-singlet components of
the $P^{0}: D^-_{s}p$ (2907 MeV), $D^-\Sigma^+$ (3058 MeV) and
$\overline{D^{0}}\Lambda$ (2981 MeV). The relative strengths of these
components depend strongly on the binding energy, as discussed above for
the deuteron-like Pentaquark. Pentaquark searches in progress in E791 \cite
{791,jl791} are based on charged particle decay modes of different
Pentaquark components: $D_{s}^-p \rightarrow \phi \pi^- p$ (B=3\%),
$D_{s}^-p \rightarrow K^{*0}K^-p$ (B=3\%), $D^{-} \Lambda \rightarrow
K^+\pi^-\pi^-\Lambda$ (B=8\%), $\overline{D^{0}}\Lambda \rightarrow
K^-\pi^+\Lambda$ (B=4\%) and $\overline{D^{0}}\Lambda \rightarrow
K^-\pi^+\pi^+\pi^-\Lambda$ (B=8\%). The indicated branching ratios are
those of the on-shell D-meson. Weak decays of virtual color singlet
substructures in bound states
are possible, $\Lambda D^0$ or $\Sigma^+ D^-$ for example,
if their masses are smaller than the $D_{s}^-$p threshold. In
other cases,  there would be strong decays through quark rearrangement
($\Sigma^+ D^-)_{bound} \rightarrow D_{s}^- + p$, and so on. Even if the
masses are smaller, the phase space favors decay to the lightest system.
The phase space factor would cause the partial width for any decay mode to
be smaller than for the on-shell decay, making the total lifetime longer.

The decay through the direct five quark [non-cluster] component can
open many additional channels; such as two-particle $\pi^-$p, K$^-$p, and
$\Xi^-$K$^+$ final states. These additional decay modes can shorten the
lifetime of the Pentaquark, which would reduce the experimental
possibilities to observe it. Such relatively simple final states are more
prone to contamination by large combinatoric backgrounds.

Consider the resonant Pentaquark possibility. Yields can be high, as one
measures the total strong decay, rather than a particular weak decay mode.
The width is the crucial parameter that determines the possibility to
observe  a resonance. Chances for observation would be good if it is of the
order of 50-100~MeV or lower, similar to widths of excited D$^*$ mesons and
widths estimated by Greenberg and Lomon \cite {lg} for the lowest lying
strangeness -1 dibaryon resonances. Our attitude is to support experimental
searches for narrow exotic Pentaquark resonances.

\section{Experimental Pentaquark Search}

An experimental program to search for the Pentaquark should include:\\
(1) Reactions likely to produce the Pentaquark, complemented by an estimate
of the\\production cross section.\\
(2) Experimental signatures that allow identification of the Pentaquark.\\
(3) Experiments in which the backgrounds are minimized. \\
These points will be further discussed in the
following subsections.

\subsection {Experimental Considerations}

All charm experiments require vertex detectors consisting of many planes of
silicon micro-strips with thousands of channels. E791 used 23 such planes.
Some of the planes are upstream of the target. These detectors allow a
high efficiency and high resolution for reconstruction of both primary
(production) vertex and secondary (decay) vertex. The position resolution
of the vertex detectors is typically better than 300 microns in the beam
direction. By measuring the yield of a particle as a function of
the separation between the two vertices, the lifetime
of the particle is obtained. Other major components of the
spectrometers are dipole magnets for momentum analysis, wire chambers for
track reconstruction, cerenkov counters for particle identification, and
Electromagnetic and Hadronic calorimeters. Muon detectors are included for
studies of leptonic decays. The invariant mass resolution for typical
charm masses in such
spectrometers is about 10 MeV. Different  spectrometers are
sensitive to different regions of Feynman-x values.

In hadronic production, the charm states produced are preponderantly charm
mesons at low x. The triggers for such experiments vary. In E791, the
requirement was to ensure an interaction in the target (using signals from
various scintillators) and a transverse energy ($E_{t}$) larger than some
threshold. The rest of the charm selection was done off-line. Increased charm
sensitivity can be achieved as in E781 \cite {cb781} by a trigger condition
that identifies a secondary vertex. A good charm trigger can produce an
enriched sample of high x charm baryons with improved reconstruction
probability because of kinematic focusing and lessened multiple scattering.
Charm2000 experiments will also require charm enhancement triggers \cite
{jeff}. The present E791 and future E781 and Charm2000 experiments \cite
{dk} complement each other in their emphasis on different x regions,
incident particle types, statistics and time schedules.

\subsection{Pentaquark Production Mechanisms}

 We consider possible mechanisms for P formation. For the central
hadron-nucleus charm production at several hundred GeV/c, the elementary
process is often associated with $q\bar{q} \rightarrow c \bar{c}$ or $gg
\rightarrow c \bar{c}$ transitions. The produced charmed quarks propagate
and form mini-jets as they lose energy. Hadronization associated with each
jet proceeds inside the nucleus, and to some extent also outside the
nucleus; depending on the transverse momentum of the jet. The propagating
charmed quarks may lose energy via gluon bremsstrahlung or through color
tube formation  in a string model, or by other mechanisms, as discussed in
ref. \cite {nied} and references therein. One may form a meson, baryon,
Pentaquark, according to the probability for the charmed quarks to join
together with appropriate quarks and antiquarks in the developing color
field. One can estimate Pentaquark production cross sections via one-step
and also two-step hadronization. All such estimates are very rough. Our aim
is to account for major ingredients in estimating the cross section, and to
give a conservative range of values. For one-step hadronization, the
$\bar{c}$ joins directly to the other quarks. The one-step is the usual
mechanism for meson and baryon formation. For two-step, the first involves
meson and baryon hadronization, while the second involves meson-baryon
coalescence.

We first consider estimates for the central production cross section
assuming a meson-baryon coalescence mechanism, expected to be the main
mechanism for production through the long-range (deuteron-like) component
of the Pentaquark wave function. We make a crude estimate relative to the
D$_{s}^-$, an anticharmed-strange meson ($\bar{c}s$). The weakly bound P
(deuteron type structure) can be produced by coalescence of a proton or a
neutron with a D$_{s}^-$, analogous to the production of a deuteron by
coalescence of a neutron and a proton. The data \cite {dbar} give roughly
10$^{-3}$ for the $\sigma(d)/\sigma(p)$ production ratio. This ratio can
also be applied to $\sigma(P)/\sigma(D_{s}^-)$ production. The reason is
that in both cases, the same mass (nucleon mass) is added to the reference
particle (proton or D$_{s}^-$), in order to form a weakly bound
deuteron-like state.

We now consider the one-step hadronization of a Pentaquark, expected to be
the main mechanism for the production through the short-range component of
the Pentaquark wave function. We rely here on an empirical formula which
reasonably describes the production cross section of a mass M hadron in
central collisions. The transverse momentum distribution at not too large
p$_t$ follows a form given as \cite {hag}:
\begin{equation}
d\sigma/dp_t^2 \sim exp(-B\sqrt{M^2+p_t^2}),
\end{equation}
where B is roughly a universal constant $\sim$ 5 - 6 (GeV)$^{-1}$. The
exponential fit has inspired speculation that particle production is
thermal, at a temperature B$^{-1}$ $\sim$ 200~MeV \cite {hag}. One can also
include a (2J+1) statistical factor to account for the spin of the hadron.
To illustrate the universality of B, we evaluate it for a few cases. For
$\Lambda_c$ and $\Xi^0$, empirical fits to data
give exp(-$bp_t^2$), with b=1.1
GeV$^{-2} $and b=2.0 GeV$^{-2}$, respectively \cite{wa89,rot}. This
corresponds to B=~5.0 GeV$^{-1}$ for $\Lambda_c$, and B=~5.3 GeV$^{-1}$ for
$\Xi^0$. For inclusive pion production, experiment gives exp(-$bp_t$) with
b =~6 GeV$^{-1}$ \cite {pi6}; and B $\sim$ b, since the pion mass is
small. Therefore, B= 5-6 GeV$^{-1}$ is valid for $\Lambda_c$, $\Xi^0$
hyperon, and pion production. We expect therefore that eq. 4 should be also
applicable to Pentaquark production. After integrating over p$_t^2$, we
estimate the ratio:
\begin{equation}
\sigma(P)/\sigma(D_{s}^-) \sim exp[-5[M(P)-M(D_{s}^-)]] \sim 10^{-2}.
\end{equation}
\noindent
For illustration, let us consider the ratio of $\Lambda_c$ to D$_{s}^-$
total production cross sections by sufficiently energetic baryon beams.
This ratio is roughly 0.23, comparing the $\Lambda_c$ cross section \cite
{wa89} with incident $\Sigma^-$ to the $D_{s}^-$ cross section \cite{ds}
with incident neutron. Eq. 5 with the masses of these particles, including
a spin statistical factor, gives about the same ratio. In applying eq. 5 to
Pentaquark production, we assume that the suppression of cross section for
the heavy P as compared to the light D$_{s}^-$ is due to the increased mass
of P. The particular one-step hadronization process is not relevant.
However, as the size of the P increases, this formula would be less and
less reliable. Cross section estimates for P production have been given
previously \cite {791,jl791}, based on other arguments, and are consistent
with the ratio given by eq. 5.

All the various reaction mechanisms described above can contribute to the
production cross section, which is estimated in the range of
$\sigma(P)/\sigma(D_{s}^-) = 10^{-3} - 10^{-2}$. In actual measurements,
the product $\sigma$ $\cdot$ B for a particular decay mode is
measured, and estimates of the P lifetime and
branching ratios may be necessary as well.

\newpage
\subsection{Pentaquark Expected Yield}

 We proceed with count rate estimates. Analysis of a part of the E791 data
(500 GeV/c $\pi^-$ beam) already yielded a preliminary upper limit
$\sigma(P^0) / \sigma(D^-_{s}) < 6\% $ for Pentaquark production \cite{sdpf}.
This was done for the $D^-_s \rightarrow \phi \pi^- $ and $P^0 \rightarrow
\phi \pi^- p$ decays assuming the same branching ratios. It was based on a
small fraction of the data and measured $D^-_s$ yield. With the full data
sample, several tens of Pentaquarks may be observed if the cross section is
in the range estimated in the previous section. For the planned E781 and
charm2000, when both use Baryon beams, we rely on previous measurements
done with similar beams. With 600 GeV/c neutrons, the D$_{s}^-$ was
measured \cite {ds} in the $D_{s}^- \rightarrow \phi \pi^-$ decay mode with
$\sigma$B =0.76 $\mu$b/N for $0.05 < x  < 0.3$, where x designates the
Feynman x-value. For Baryon beams the cross section should be proportional
to $(1 - x)^{n}$, with n between 4.5 and 5.5, based on the WA89 experiment
\cite {wa89} with a 300 GeV/c $\Sigma^-$ beam. These data and x-dependence
correspond to $\sigma$ $\cdot$ B values for the whole range of $x >$ 0 of
roughly 1.~$\mu$b/nucleon.
With the $\sigma(P)/\sigma(D_{s}^-)$ factors given above, we estimate
$\sigma \cdot$ B = 1 - 10 nb/N, for each of P$^0$ and P$^-$. For E781,
scheduled for 1996, the experimental conditions should allow reconstructed
Pentaquark events at a rate of roughly 200 events/nb. These expectations
are based on a contribution to this workshop by J. Russ \cite{cb781}, which
cites an expected yield of 2300 charm events/nb of cross section for 100\%
efficiency. The efficiencies include a  tracking efficiency of 96\% per
track, a trigger efficiency averaged over x of roughly 18\%, and a signal
reconstruction efficiency of roughly 50\%. We therefore assume an overall
average Pentaquark reconstruction efficiency of $\varepsilon \simeq 8\%$.
We then estimate an expected yield of N(P$^0$)= 200 - 2000 in E781. If we
assume a rate of 2000 events/nb for charm2000, the Pentaquark yield may
reach the 2000 - 20,000 range. These projections depend critically on the
value used for the D$_s$ production cross section. We note that the
value quoted in \cite {ds} is exceptionally large.

 It is still possible that different mechanisms for charm production
contribute in different x regimes. For example, there is evidence for
leading production of charmed hadrons in WA89 and FNAL E769 \cite{769},
which suggests  diffractive contributions. For charm2000, one could study
\cite {mm} the pair diffractive production reaction $p+N \rightarrow (P^0
D_{s}^+)+N$, with possible $D_{s}^+$ tag or without such tag. For the
diffractive pair production cross section, one can compare to the
diffractive cross section for the reaction $p + N \rightarrow (\Lambda K^+)
+ N$ at 70 GeV; about 4 $\mu b$ after subtraction of isobar contributions
\cite {lmk}. Estimates are needed but are not available for the cross
section ratio $\sigma(P^0 D_{s}^+)/\sigma(\Lambda K^+)$. For the ratio of
10$^{-3}$, with B = 3\%, one would obtain around 240 reconstructed $P^0$
baryons with charm2000. There is the $D_{s}^+$ tag possibility for this
process. The efficiency for tagged versus untagged events is reduced, but
tagging may improve the signal to background ratio.

\subsection{Pentaquark Decay Signatures}

(1) Mass and Width  and Decay Modes:

   Searches for the Pentaquark are easiest via modes having all final decay
particles charged. With all charged particles detected, the invariant mass
of the system can be determined with high resolution. One signature of the
Pentaquark is a peak in the invariant mass spectrum somewhat lower than
2907 MeV if the system is bound, and above if it is a resonance. The
position of the peak should be the same for several decay modes. It's width
should be determined by the experimental resolution if it is bound, and
broader if it is a resonance.

The selection of the decay modes to be studied is made primarily by
considering detection efficiency and expected branching ratios. Since the
$D_{s}^-p$ system is the lightest it is expected to be preferred from phase
space arguments. Also, two of it's decay modes have four charged particles
in the final state (e.g. $K^+K^-\pi^-p$ : $\phi \rightarrow K^+ K^-$,
${K^{*}} \rightarrow K^+ \pi^-$). We describe how this signature is
implemented. First, two distinct vertices are identified: a production
vertex and a decay vertex. From the decay vertex, four tracks are
identified and associated with $K^+K^-\pi^-p$. By reconstructing the
invariant mass of the $K^+K^-$ pair, one can require only $\phi$ mass
events. One then reconstructs the invariant mass of all four particles. If
there is a peak in the resulting spectrum, it will be one of the
identifying characteristics of the Pentaquark. One can also study a strong
decay into $D_{s}^-p$, if the P is a resonance. For this strong decay, the
proton and D$^-_{s}$ come from primary vertex, and the D$^-_{s}$ decay
forms the secondary vertex. Both weak and strong decay modes coming from
the $D_{s}^-p$ and the $\overline{D^{0}}\Lambda$ components of the P are
currently being studied in E791.

(2) One General Signature - A Spectator Baryon:

We first note a striking signature for Pentaquark decay which may be useful
for discrimination against background. This signature is predicted by both
of two very different Pentaquark models (1) a loosely-bound $D_{s}^-p$
deuteron-like state and (2) a strongly-bound five-quark state. Both models
predict decay modes into a baryon and two or more mesons, in which the
three quarks in the baryon are spectators in the decay process and remain
in the final state with a low momentum which is just the fermi momentum of
the initial bound state.

That the baryon is a spectator is obvious in the deuteron model, in which
the decay is described as an off-shell $D_{s}^-$ decaying with a nucleon
spectator. In the five-quark model, a similar situation arises in the
commonly used spectator model with factorization. Here, the charmed
antiquark decays into a strange antiquark by emission of a $W^-$ which then
creates a quark-antiquark, which hadronizes into mesons. The strange
antiquark combines with one of the four spectator quarks to form one or
more mesons, while the three remaining spectator quarks combine into a
baryon.

In both cases, it seems that the final state should show a low-momentum
baryon in the center-of-mass system of the Pentaquark and the invariant
mass spectrum of the remaining mesons peaked at the high end near the
kinematic limit. Thus in the particular cases of the $p \phi \pi^-$,
$K^{*o}K^-p$ and $\Lambda K^+ \pi^-$ decay modes, the $\phi \pi^-$, $K^{*o}
K^-$ and $K^+ \pi^-$ invariant mass distributions respectively should show
this peaking near the kinematic limit.

Note that in the particular case of the $p \phi \pi^-$ decay mode, a low
momentum proton in the center of mass system means that the $\pi^-$ and
$\phi$ are back to back with the same momentum and therefore that the pion
carries off most of the available energy. Thus one might reduce background
with a cut that eliminates all pions with low momentum in the center of
mass.

(3)  Some Model-Dependent Branching Ratio Predictions:

The $\phi\pi^-p$ decay mode is the most convenient for a search, since the
$\phi$ signal is so striking. We now examine the lowest order predictions from
the two extreme models for the branching ratios of other modes relative to
$\phi\pi^-p$.

In experiments sensitive only to charged particles the $\phi\pi^-p$ decay mode
is observed in the four-prong final state $ K^+ K^- \pi^- p$. The $K^{*o}K^-p$
decay mode is also observable in this same four prong final state. The $K^{*o}
K^-p$ decay mode arises naturally in the deuteron model, since the $K^{*o}
K^-$ decay is observed for $D_{s}^-$ decays with a comparable branching ratio
to
$\phi \pi^-$. In this model, the ratio of the two decays is predicted from
observed $D_{s}^-$ decay branching ratios with phase space corrections.
However,
the $K^{*o} K^-p$ decay mode does not occur in the five quark spectator model,
where the spectator strange quark can only combine with the $\bar s$ produced
by the charm decay to make a $\phi$ or with two spectator nonstrange quarks to
make a hyperon. Comparing the two decays thus tests the decay model.

The $K\pi\Lambda$ and $K^*\pi\Lambda$ decay modes arise naturally in the
five quark spectator model. However, they should not be expected in a very
weakly bound deuteron model with mainly a $D_{s}^-$p structure. In that
case, the $D_{s}^-$ decays into mesons containing one strange
quark-antiquark pair and the baryon spectator has no strangeness.

(4) Angular Momentum Constraints and Angular Distributions
for P Decays:

We can give a model-independent prediction. The Pentquark has spin 1/2 and this
total angular momentum is conserved in the decay. Since the production process
is a strong interaction which conserves parity, the Pentaquark will not be
produced with longitudinal polarization. Its polarization in the beam direction
must also vanish. Therefore, the angular distribution in the center-of-mass
system of the Pentaquark must therefore be isotropic for the momentum of any
final state particle in any decay mode with respect to either the incident beam
direction or the direction of the total momentum of the Pentaquark. The
background does not necessarily have these constraints.

We also give a model-dependent prediction. We first consider the deuteron
model. The $D_{s}^-$ has spin zero, and spin is preserved in the decay.
Thus, in the center of mass frame of all the $D_{s}^-$ decay products, the
angle between the proton momentum and the momentum of any particle emitted
in the $D_{s}^-$ decay must have an isotropic angular distribution.

A further prediction is obtainable for the case of a vector-pseudoscalar
decay mode of the $D_{s}^-$; e.g. $\phi\pi^-$ or $K^{*0} K^-$. The vector
meson must be emitted with zero helicity in the rest frame of the
$D_{s}^-$. The zero helicity can be seen in the  $\phi \pi$ decay by
measuring the angle $\theta_{K \pi}$ between the kaon momenta in the $\phi$
rest frame and the pion momentum. The prediction is to have a $\cos^2
\theta_{K \pi}$ distribution. By contrast, the five-quark model for the
Pentaquark favors helicity one over helicity zero for the vector meson by
just the 2:1 ratio needed to give an isotropic distribution in $\theta_{K
\pi}$. Here again the background does not necessarily have these
constraints.

\newpage
\subsection{Reducing Background}

There is much background from central interactions. When low x production
is studied, the momenta of P$^0$ decay products are also lower.
As a result, the background rate
increases faster than the charm signal.
 It is known \cite {lmk} that the combinatoric background in inclusive
processes is significantly reduced in the fragmentation region ($x \geq
0.6$).
The produced particles and the decay fragments from the $P$, especially for
high-x production, are all focused in a forward cone in the laboratory
system. One has therefore a good efficiency for detecting all particles in
the final state.
The diffractive pair production reactions with low combinatoric
background also contribute in this high x region. One would expect more
favorable background conditions at high x for the identification of
resonance P baryon states.

High quality particle identification (PID) for the largest possible energy
range of the outgoing particles is important for reducing backgrounds
associated with incorrect  identification of tracks. This is available in
E781, for example, via ring imaging Cerenkov (RICH) and transition radiation
detector (TRD) PID systems. The separation of vertices is very important
also for reducing the combinatoric backgrounds, as the majority of
particles come from the primary vertex. These and other
experimental techniques  to reduce backgrounds are described in more detail
in the contribution of J. Russ \cite {cb781}.

\section{Heavy Baryons with Hidden Charm}

 In recent years, several candidates were reported for baryon
states  with unusual properties (narrow decay widths, large branching
ratios for the decays with strange particles). There are candidates for
cryptoexotic baryons with hidden strangeness
$B_{\phi}=\mid$~$qqqs\bar{s}\!>$ ($q=u$ or $d$ quarks) \cite {Cq}. Although the
existence of such a baryon is not yet confirmed \cite {bal}, the
suggestions raise the question of the possible existence of heavy
cryptoexotic baryons with hidden charm $B_{\psi}=\mid$~$qqqc\bar{c}\!>$. If
M$(B_{\psi})<$M$(\eta_{c})+$M($p$) $\simeq$ 3.9~GeV, the $B_{\psi}$ decays
would be OZI suppressed and the width of this cryptoexotic  baryon would be
quite narrow ($\leq$1~MeV). To search for such $B_{\psi}$ states, it was
proposed \cite {D} to use the diffractive production reaction $p+N
\rightarrow B_{\psi}^+ + N$; with possible decays of $B_{\psi}$ baryons
$B_{\psi}^+ \rightarrow p + (J/\psi)_{virt} \rightarrow p+ (l^+ l^-)$ or
$B_{\psi} \rightarrow p+(\eta_c)_{virt} \rightarrow p +
(K^+K^-\pi^+\pi^-;2\pi^+2\pi^-; K\bar{K}\pi;\eta\pi\pi)$. The $\sigma \cdot
B$ was estimated as roughly 1.5 ~nb~ \cite{D}. Assuming the expected
Charm2000 efficiency of 2000 events/nb would hold for these events
too, this would correspond to the detection of roughly 3000 events.

    If M$(B_{\psi})>$4.3~GeV, there would be OZI allowed decays $B_{\psi}^+
\rightarrow p + J/\psi; \Lambda_C + D^0$, etc. Because of a complicated
internal color structure of this baryon (see Introduction), one expects a
narrow decay width ($\leq$~100~MeV). Such resonance states may be
observable in diffractive production reactions.

\newpage
\section{Conclusions}

We described the expected properties of Pentaquarks. Possibilities for
enhancing the signal over background in Pentaquark searches were
investigated. General model-independent predictions were presented as well
as those from two models: a loosely bound $D_{s}^-N$ "deuteron" and a
strongly-bound five-quark model. While the current E791 may have marginal
sensitivity, future experiments with more than $10^5$ reconstructed charmed
baryon events should have sensitivity to determine whether or not the
Pentaquark exists.

\section{Acknowledgements}

Thanks are due to J. Appel, P. Cooper, L. Frankfurt, S. Gavin, D. Kaplan,
M. A. Kubantsev, J. Lach, J. Lichtenstadt, S. May-Tal Beck, S. Nussinov, J.
Russ, and B. Svititsky for stimulating discussions. This work was supported
in part by the U.S.-Israel Binational Science Foundation, (B.S.F.)
Jerusalem, Israel.

E-mail addresses of authors are: murray@tauphy.tau.ac.il,
ashery@tauphy.tau.ac.il,\\
lgl@mx.ihep.su, ftlipkin@weizmann.weizmann.ac.il

%

\end{document}